\documentclass[12pt,titlepage]{article}
\usepackage{geometry,graphicx,epsfig}
\usepackage{chicago}

\usepackage{natbib}
\begin{document}
\title{Use and Communication of Probabilistic Forecasts}
\author{Adrian E. Raftery \\
University of Washington}
\date{\today}
\maketitle 

\begin{abstract}
Probabilistic forecasts are becoming more and more available.
How should they be used and communicated? What are the obstacles
to their use in practice? I review experience with five problems
where probabilistic forecasting played an important role.
This leads me to identify five types of potential users:
Low Stakes Users, who don't need probabilistic forecasts;
General Assessors, who need an overall idea of the uncertainty in the
forecast; Change Assessors, who need to know if a change is out of line
with expectatations; Risk Avoiders, who wish to limit the risk of an
adverse outcome; and Decision Theorists,
who quantify their loss function and
perform the decision-theoretic calculations. 
This suggests that it is important to interact with users and 
to consider their goals.  The cognitive research tells us that calibration 
is important for trust in probability forecasts, and that it is important 
to match the verbal expression with the task. The cognitive load should be 
minimized, reducing the probabilistic forecast 
to a single percentile if appropriate.  Probabilities of adverse events and 
percentiles of the predictive distribution of quantities of interest seem 
often to be the best way to summarize probabilistic forecasts.  
Formal decision theory has an important role, but in a limited range 
of applications.
\end{abstract}

\pagenumbering{roman}
\tableofcontents
\listoffigures
\listoftables
\newpage

\baselineskip=18pt

\pagenumbering{arabic}
\section{Introduction}
Much progress has been made over the past few decades in the development
of methods for probabilistic forecasting, and probabilistic forecasts
are now routinely used in several disciplines. These include finance,
where trading decisions are made based on predictive distributions of
assets, often using automated computer trading programs.
In marketing, predictive distributions of future sales
and inventory are commonly made using statistical models such as 
ARIMA models \citep*{BoxJenkins2008}, and used as the basis for
stocking and other decisions.

However, in other areas the development of probabilistic forecasting
methods is more recent, and use of these methods in practice is at an
earlier stage. How should probabilistic forecasts be used and communicated?
What are the obstacles to their use in practice? Can these be overcome?
Can they be presented in ways that make them more useful to 
possibly sceptical users?

Communicating uncertainty is inherently a challenging problem.
\citet{Kahneman2011} identified people's resistance to uncertainty as 
\begin{quote}
``a puzzling limitation of our mind: our excessive confidence in what we
believe we know, and our apparent inability to acknowledge the full extent
of our ignorance and the uncertainty of the world we live in. We are prone
to overestimate how much we understand the world and to underestimate
the role of chance in events. Overconfidence is fed by the illusory
certainty of hindsight.''
\end{quote}

There are various possible explanations for this. 
One is that people's cognitive bandwidth is limited, and uncertainty
information increases cognitive load. For example, adding
a range to a point or ``best'' forecast triples the cognitive load.

A more fundamental explanation is proposed, again by \citet{Kahneman2011}:
\begin{quote}
``An unbiased appreciation of uncertainty is a cornerstone
of rationality, but it is not what people and organizations want.
Extreme uncertainty is paralyzing under dangerous circumstances, and the
admission that one is merely guessing is especially unacceptable when the
stakes are high. Acting on pretended knowledge is often the preferred 
solution.''
\end{quote}

A related possible explanation arises when forecasters and decision-makers
are different people, as is often the case in policy-making contexts.
Then the decision-maker may wish to push the responsibility for
the decision onto the forecaster, and when the forecasters provides a range
or a probabilistic forecast, this is harder to do than when a single 
number is given. If things go wrong, it's easier to blame the forecaster
who gave an incorrect forecast.

In this article, I will describe experience with probabilistic forecasting
in five different contexts and try to draw some conclusions.
These will lead me to identify five types of potential users of
probabilistic forecasts: Low Stakes Users, General Assessors, 
Change Assessors, Risk Avoiders, and Decision Theorists.
Each may have different needs.

Some suggestions are that it is important to interact with users and consider
their goals; ways of doing this include meetings and web surveys.
This is a cognitive problem as well as a statistical one.
The cognitive research tells us that calibration is important for
trust in probability forecasts, and that it is important to match the
verbal expression with the goal. The cognitive load should be minimized
to the extent possible, even reducing the probabilistic forecast
to a single number if appropriate.  Probabilities of adverse events
and percentiles of the predictive distribution of quantities of interest
seem often to be the best way to summarize probabilistic forecasts.

Formal decision theory has an important role in a limited range of applications,
particularly when users are aware of their loss functions, and when
there is agreement on the loss function to use. 
This arises most clearly when costs and losses are measured in monetary terms.
Decision theory is also useful in research on the use of probabilistic
forecasts, to analyze different possible decision rules.

This article is organized as follows. In the following sections I will
describe experience with five problems where probabilistic forecasting 
played an important role:
setting aboriginal whaling quotas, probabilistic weather forecasting,
projecting the worldwide HIV/AIDS epidemic, probabilistic population
projections for the UN, and deciding on the number of funded graduate students to admit.
I will then discuss what conclusions can be drawn from this experience.

\section{Setting Aboriginal Bowhead Whaling Quotas}
\label{sect-iwc}
\begin{figure}
\begin{center}
\begin{tabular}{cc}
\includegraphics[height=0.22\textheight]{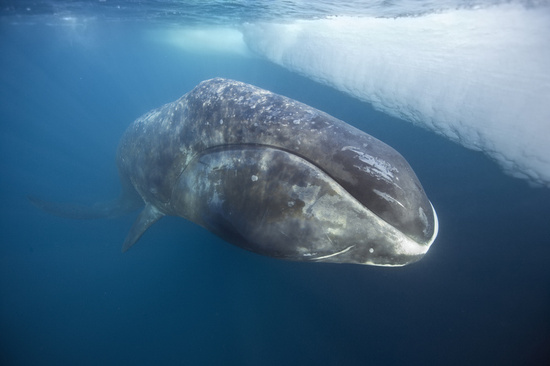} &
\includegraphics[height=0.22\textheight]{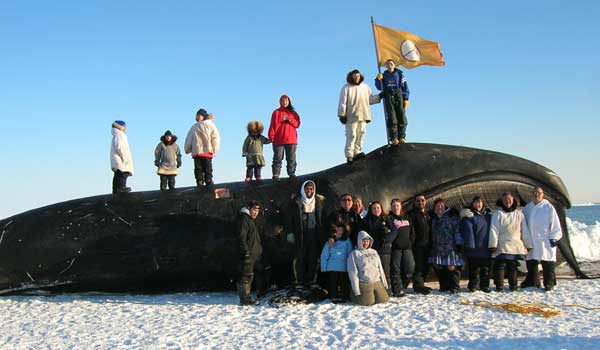} 
\end{tabular}
\end{center}
\caption{\label{fig-bowhead} Left: Bowhead whale, {\it Balaena mysticetus}.
Right: Community celebration after Inuit bowhead whale hunt.}
\end{figure}

For centuries, the Western Arctic stock of bowhead whales, 
{\it Balaena mysticetus}, off the coasts of Alaska and Siberia,
has been the object of small-scale subsistence
hunting by the Inuit, or Eskimo, peoples of the area,
for whom it is vital both nutritionally and culturally;
see Figure \ref{fig-bowhead}.
The stock was severely depleted by commercial whaling by Yankee and
European whalers in the late 19th and early 20th centuries.
Commercial whaling of bowhead whales (although not other whale species)
effectively ended around 1915, and the species was first protected
legally from commercial whaling from 1931 by the League of Nations Convention,
and then by the International Whaling Commission (IWC), founded in 1946.

This left the question of whether and how to regulate
aboriginal whaling by the Inuit.
It was generally recognized that it would be unfair to penalize the
Inuit people for a problem that was not of their making, since they
had been whaling sustainably for centuries, and that to ban 
aboriginal whaling would damage their livelihood and culture.
This led to a tension between two conflicting goals: on the one hand, to protect
the whale stock and allow it to recover to its pre-commercial whaling levels,
and on the other hand
to satisfy the subsistence and cultural needs of the Inuit people.

The IWC's solution was to allow continued limited aboriginal subsistence
whaling, but with a quota to be set at a level low enough to allow the
stock to recover. A key quantity for setting the quota in a
given future year is the replacement yield (RY) in that year, namely
the greatest number of whales that could be taken without
the population decreasing. This is unknown and is subject
to considerable uncertainty. Because it is important that the
quota not exceed this unknown value, a conservative value
or ``lower bound'' is sought, which should take account of all
nonnegligible sources of uncertainty.

The future RY has traditionally been forecast using a 
deterministic population dynamics model, in which births are added
and natural deaths and kills are subtracted. This requires 
as inputs age-specific fertility and natural mortality rates,
and outputs the population for each future year, broken down by age 
and sex. The inputs are unknown and subject to considerable uncertainty.

Until 1991, the lower bound was set by doing several runs of the model
with different scenarios or variants, consisting of 
combinations of ``central,'' ``high,'' and ``low'' values
of the inputs. The range of values of RY output was then treated as a rough
prediction interval. In 1991, however, the IWC Scientific Committee
rejected this approach as statistically invalid, noting that it
had no probabilistic interpretation and could lead to, for example,
decisions that were riskier than they seemed. They recommended
that statistically principled methods be developed.

\begin{figure}
\begin{center}
\includegraphics[height=0.25\textheight]{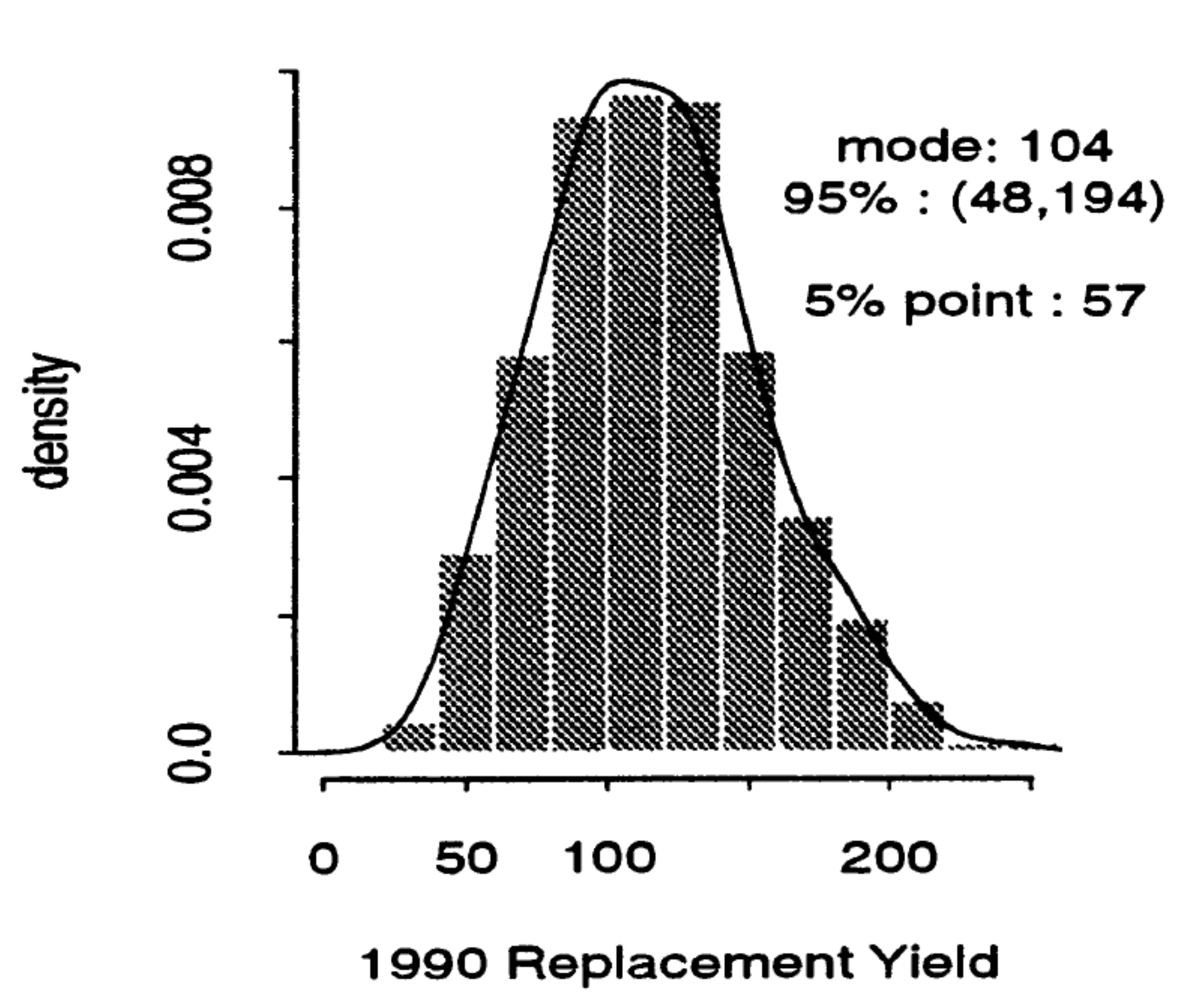}
\end{center}
\caption[Posterior Predictive Distribution of the
1990 Replacement Yield of Bowhead Whales, obtained by Bayesian Melding]
{\label{fig-RY} Posterior Predictive Distribution of the 
1990 Replacement Yield of Bowhead Whales, obtained by Bayesian Melding.
{\it Source:} Raftery, Givens and Zeh (1995).}
\end{figure}

In response to this, we developed the Bayesian melding method for
making inference about outputs from the population dynamics model,
taking account of all known substantial uncertainties about the inputs
\citep*{rgz1992,rgz1995,PooleRaftery2000}.
This yielded a posterior predictive distribution for RY for future
years; an example is shown in Figure \ref{fig-RY}.

Once this was available, the IWC Scientific Committee recommended that
the 5th percentile of this distribution be taken as a precautionary
lower bound on RY, and thus as an upper bound on the allowed hunting quota.
The recommendation was accepted by the Commission itself
(consisting mostly of politicians and senior civil servants, such as
Fisheries Ministers and officials from the then 40 IWC member countries).
Taking account of this lower bound, the desire to allow a margin for
future recovery of the stock, and the Inuit subsistence and cultural needs,
the Commission set a quota slightly below the lower bound. 

This approach was used successfully for the following ten years.
Over that period, the bowhead whale stock prospered,
indeed increasing substantially, while the Inuit whale hunt continued and
the related Inuit culture was preserved. The basic statistical ideas have
since been used for other wildlife management problems 
\citep{Powell&2006,Brandon&2007,Falk&2010}.

The 5th percentile of the posterior predictive
distribution effectively became the ``point forecast'' for this problem.
To calculate it, it was necessary to compute the full posterior distribution.
But once the 5th percentile had been calculated and agreed as valid by
the IWC Scientific Committee, most of the policy-making attention focused on it,
and the rest of the distribution (including measures of its center such as 
the median or mode) was largely ignored. Thus the cognitive load was no larger
than for a single ``best'' forecast. 

Also, the responsibility for 
making a single best forecast had been met by the forecasters
(in this case the IWC Scientific Committee) --- only in this case
it was a lower bound rather than a predictive median or mode, or a deterministic
point forecast. Probabilistic forecasts were important in this application
because the first priority was to limit the risk of an adverse outcome,
namely a decrease of the whale stock.

Note that formal decision theory was not used in this problem.
The IWC Scientific Committee has considered using formal decision theory
for such problems, but in general has not done so, because they considered
that reaching agreement on the relative costs involved was not feasible.
For example, what is the ratio of the cost to the stock of killing a whale 
to the benefit to the Inuit community? Consensus on the answers to 
questions like these would be hard to achieve
\citep{PuntDonovan2007,Cooke&2009,Cooke&2012}.

 Instead, the preferred approach
was to set the quota so that the risk of the stock decreasing as a result
would be no more than 5\%, and this eventually commanded broad agreement,
even in a body where debates have often been contentious because of the
environmental sensitivities associated with whaling.

\section{Probabilistic Weather Forecasting}
\label{sect-pwf}

\subsection{Methods and Probcast Website}
\label{sect-pwf.probcast}
Probabilistic weather forecasts consist of predictive probability 
distributions of future weather quantities.
In particular they yield probabilities of future adverse weather
events, such as freezing temperatures, high rainfall or wind storms.
Since 1992, probabilistic weather forecasts have been produced
by major weather forecasting agencies using ensembles of 
deterministic numerical weather predictions \citep{GneitingRaftery2005}.
However, these have been little used as the basis for public forecasts,
because they are typically poorly calibrated.

In response to this situation, methods for postprocessing 
ensembles to produce calibrated probabilistic weather forecasts 
have been developed, based on statistical methods, including ensemble 
Bayesian model averaging \citep{Raftery&2005} and
ensemble model output statistics (EMOS) \citep{Gneiting&2005}.
In addition to temperature, methods were developed for 
precipitation \citep{Sloughter&2007}, wind speeds \citep{Sloughter&2010},
wind directions \citep{Bao&2010}, wind vectors \citep{Sloughter&2013},
 and visibility \citep{ChmieleckiRaftery2011}.

\begin{figure}
\begin{center}
\includegraphics[width=\textwidth]{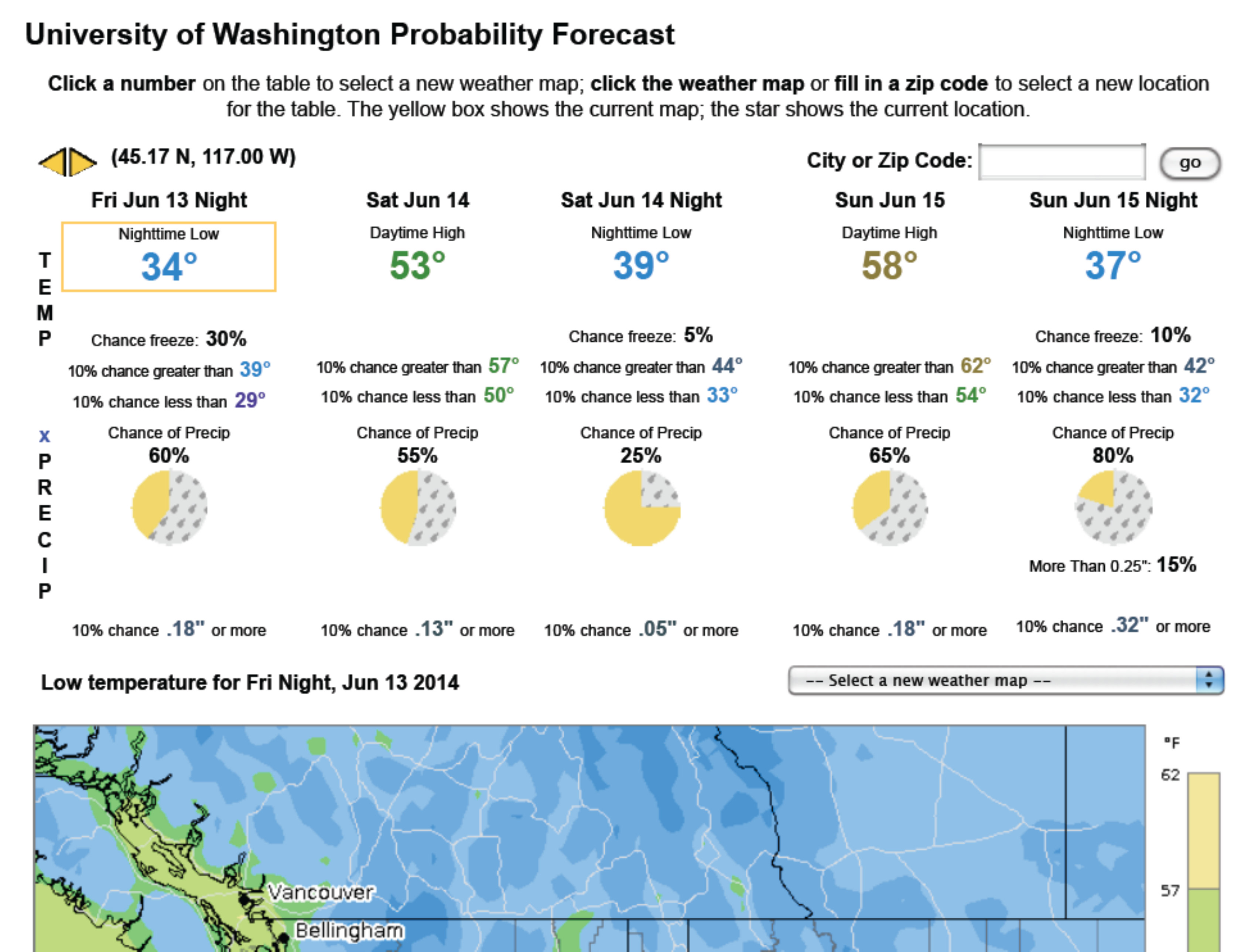}
\end{center}
\caption{\label{fig-probcast} Screenshot of the Probabilistic 
Weather Forecasting Website at {\tt www.probcast.com}.}
\end{figure}

Based on these forecasts, we set up a prototype real-time probabilistic
weather forecasting website for the general public in the North American
Pacific Northwest, at {\tt www.probcast.com} \citep{Mass&2009}; 
see Figure \ref{fig-probcast}. Its design and content
were based on extensive cognitive experiments and ethnographic studies
of forecasters and end-users 
\citep{JoslynJones2008,Joslyn&2008,Nadav-Greenberg&2008,JoslynNadav-Greenberg2009,JoslynSavelli2010}.

The website contains three kinds of information. First are percentiles
of decision-critical weather quantities, namely temperature 
and the amount of precipitation. The 10th, 50th and 90th percentiles of
future temperature are given. For precipitation the (upper) 90th
percentile is given. 

The second kind of information consists of probabilities
of adverse weather events of common interest, namely freezing temperatures,
precipitation (defined as more than 0.01 inches in the 12-hour period
of the forecast), heavy precipitation (defined as more than 0.25 inches),
and very heavy precipitation (defined as more than 1 inch).
When the probabilities are below 5\%, these fields are left blank.
The third kind of information consists of maps of any of the 
percentiles or probabilities in the upper part of the web page,
showing how they vary over the spatial domain.

\begin{figure}
\begin{center}
\includegraphics[width=\textwidth]{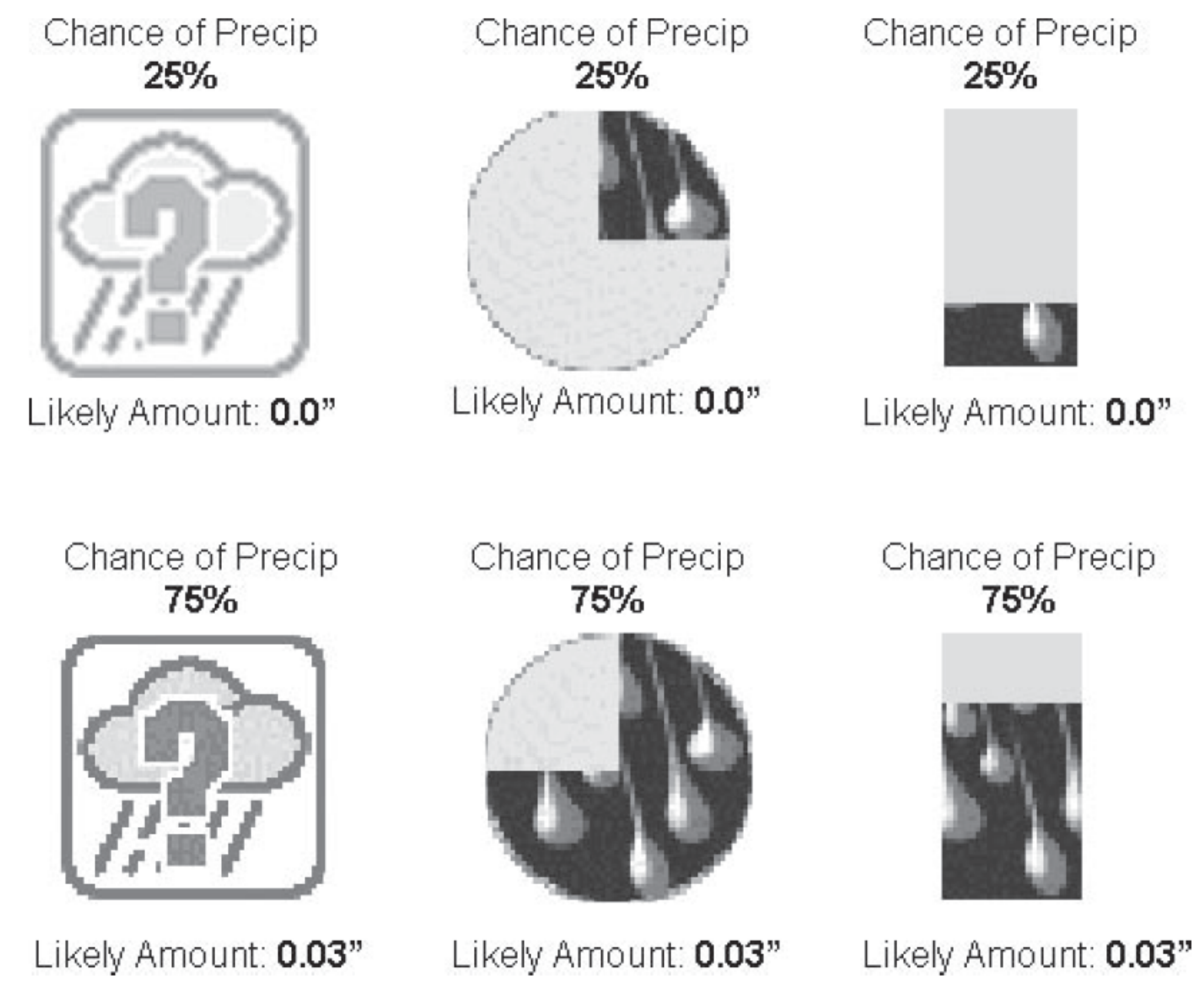}
\end{center}
\caption[Icons used in cognitive experiments to
compare the relative effectiveness of different icons for
probability of precipitation]
{\label{fig-icons} Icons used in cognitive experiments to 
compare the relative effectiveness of different icons for 
probability of precipitation: question mark icon (where the icon is
darker when the probability is higher), pie icon and bar icon.
{\it Source:} \protect\citet{JoslynNadav-Greenberg2009}.}
\end{figure}

The kinds of display used were chosen on the basis of cognitive experiments.
For example, to choose the icon representing probability of precipitation
seen in Figure \ref{fig-probcast}, cognitive experiments were carried
out to compare the relative effectiveness of several kinds of icon
\citep{JoslynNadav-Greenberg2009}.  Three of the icons are shown in 
Figure \ref{fig-icons}: a question mark icon, a pie icon, and a bar icon.
In the question mark icon, higher probability of precipitation is
represented by darker colors. The pie icon produced the fewest
misunderstandings among study participants and so was used on the
Probcast website.

On the Probcast website we gave the 10th and 90th percentiles of
temperature, corresponding to an 80\% prediction interval. 
There is a trade-off in choosing the default probability levels
to display: larger intervals (e.g. the 95\% interval) contain a 
higher proportion of actual outcomes, but they are also much wider,
and hence may be judged less useful. In the event, we received almost
no requests for higher probability level intervals, and so we
stuck with the 80\% intervals. It would of course be possible to 
display multiple probability levels, but this would add to the 
cognitive load and so make the website harder to use.

\subsection{Cognitive Findings}
\label{sect-pwf.cognitive}
An important part of the probabilistic weather forecasting project
consisted of carrying out cognitive experiments to determine how
best to convey the uncertainty information.
There is a long tradition in psychology of assessing people's 
understanding of probability and uncertainty by offering them
simple gambles \citep{KahnemanTversky1984}, but less research
on how best to communicate uncertainty about complex real-life outcomes.

Calibration of the probability forecast (e.g. 80\% prediction 
intervals contain the truth 80\% of the time on average) is
an important requirement for probabilistic forecasts
\citep{Gneiting&2007}. One series of experiments showed that
providing calibrated probability forecasts improve weather-related
decision-making and increases trust in the forecast
\citep{Joslyn&2007,Nadav-GreenbergJoslyn2009,JoslynLeclerc2012}.
This is good news for probabilistic forecasting, showing that
ordinary people can understand and use probabilities to improve
their decision-making.

\citet{JoslynSavelli2010} found that users of standard (deterministic)
weather forecasts have well-formed uncertainty expectations,
suggesting that they are prepared to understand explicit
uncertainty forecasts for a wide range of parameters.
Indeed, explicit uncertainty estimates may be necessary to overcome 
some of the anticipated forecast biases that may be affecting the usefulness of
existing weather forecasts. Despite the fact that these bias expectations 
are largely unjustified, they could lead to adjustment
of forecasts that could in turn have serious negative consequences for users, 
especially with respect to extreme weather warnings.

\citet{Joslyn&2008} reported on a series of experiments to investigate
the effects of various aspects of how probability forecasts are presented:
framing (positive versus negative), format (frequency versus probability),
probability (low versus high), and compatibility between the 
presentation and the decision task. They showed that the key factor
is the match between the verbal expression and the task goal.
The other three factors (framing, format and probability) made
a much smaller difference.

In one experiment, people were asked to decide whether or not to post
a wind advisory for winds above 20 knots, and were given probability
information. When people were told the probability that wind speed
would be above 20 knots, they made few errors. However, when they
were told the probability that wind speed would be below 20 knots,
they made far more errors, even though the information is mathematically
equivalent. This indicates that when the verbal
expression and the task were mismatched, more errors were made.

Another series of experiments was carried out to assess whether 
it was better to present probability forecasts in terms of
probability (e.g. 10\% chance) or frequency (e.g. 1 time in 10).
It has been argued that uncertainty presented as frequency
is easier for people to understand 
\citep{Fiedler1988,HertwigGigerenzer1999}.
However, \citet{JoslynNichols2009} found that people better understood
the forecast when it was presented in probability format rather
than a frequency format, in contrast with the earlier research. 
This is more good news for probabilistic forecasting, 
indicating again that ordinary people can understand probabilities.

\subsection{Assessment}
Overall, the Probcast website has been reasonably successful, attracting
about two million unique visits since it was set up in 2005
\citep{Jones2011}.
Public probabilistic weather forecasting (beyond probability of precipitation,
which has been issued by the U.S.~National Weather Service for about 40 years)
is now being considered and evaluated by several national and other
weather agencies, and Probcast provides both a methodology for
producing calibrated probabilistic forecasts and a model of how they might 
be communicated to the public. It has also been cited by \citet{NRC2006} 
as a possible model for communication of uncertainty in weather forecasting.

While specialists sometimes argue that the public doesn't understand
probabilities and so that there's little point in issuing probabilistic
forecasts, the research results from the Probcast project suggest
otherwise. The cognitive results indicate that users are ready
for explicit uncertainty statements in forecasts, and that
including them can improve decision-making and increase trust in the forecast.
The fairly wide public use of the Probcast website, in spite of its
lack of substantial institutional backing and its narrow geographic range
(the North American Pacific Northwest), suggest that the public is
ready for probabilistic forecasts on a broader scale, although of
course only a portion of the public would actively use them
(notably those with higher-stakes weather-related decisions to make).

The cognitive experiments carried out as part of our project by
Susan Joslyn's research group at the University of Washington
suggest that probabilities of particular
adverse weather events (e.g. freezing temperatures, precipitation,
heavy precipitation, high winds), and percentiles (10th, 50th, 90th)
of the predictive distribution of continuous weather quantities
of interest (e.g. temperature, amount of precipitation, wind speed)
are useful quantities to provide to users 
\citep{SavelliJoslyn2013}.
The work suggests that both are understandable to people and that they
make better decisions when they have this information.

A common prescription is that probabilities should be used in decision-making
using decision theory \citep{vonNeumannMorgenstern1947}.
This says that each possible outcome imposes a loss on the decision-maker,
and that the decision made should minimize the expected overall loss.
In this case the expectation would be taken over possible future weather
outcomes, and the losses might relate, for example, to the costs
of issuing a high wind warning if no high winds occur, and to
the damage that high winds would cause in the absence of a warning.
This seems to be a very useful framework when the utilities associated
with different outcomes can be quantified on the same scale,
typically money. The clearest weather example that I know of is
decision-making by wind energy companies that have to bid for contracts
to provide specified amounts of energy at given prices and with specified
penalties for failing to fulfil the contract, in the presence of
great uncertainty about future wind speeds \citep{Pinson&2007}.

However, we did not incorporate decision-theoretic concepts explicitly
into the Probcast website. It seems that most people are unaware of 
their utility functions, and may even be unwilling to specify them
when the losses involved are on different scales (e.g. money versus
possible loss of life). Thus people may find it easier to use probabilistic
forecasts to make decisions that limit the risk of adverse outcomes to
acceptable levels, rather than carrying out a full decision-theoretic analysis.

Nevertheless, \citep{JoslynLeclerc2012} showed that when costs and benefits
are on the same scale (e.g. money), while people do not match the 
optimal decision-making standard, they are closer to it when they have
probabilistic information. 
\citep{JoslynLeclerc2012} also found that if people were given
decision advice based on optimal decision-theoretic calculations,
they followed the advice only if they were also given the probabilities.

\section{Projecting the HIV/AIDS Epidemic}
\label{sect-unaids}
The Joint United Nations Programme on HIV/AIDS (UNAIDS)
publishes updated estimates and projections of the
number of people living with HIV/AIDS in the countries with
generalized epidemics every two years.  Generalized epidemics
are defined by overall prevalence being above 1\% and the epidemic
not being confined to particular subgroups; there are
about 38 such countries \citep{Ghys&2004}.
UNAIDS projections are typically provided for no more than five
years into the future.
As part of this, statements of uncertainty are also provided. 

This exercise has two main goals. 
The first is to develop estimation and projection methods and software
for use by country health officials for planning, for example
to meet future medication needs. There, statements of uncertainty
may be used, for example, for determining the amount of medication needed
to be reasonably sure of having enough to meet the need;
this would correspond to an upper percentile of the predictive distribution.

\begin{figure}
\begin{center}
\begin{tabular}{cc}
\includegraphics[width=0.45\textwidth]{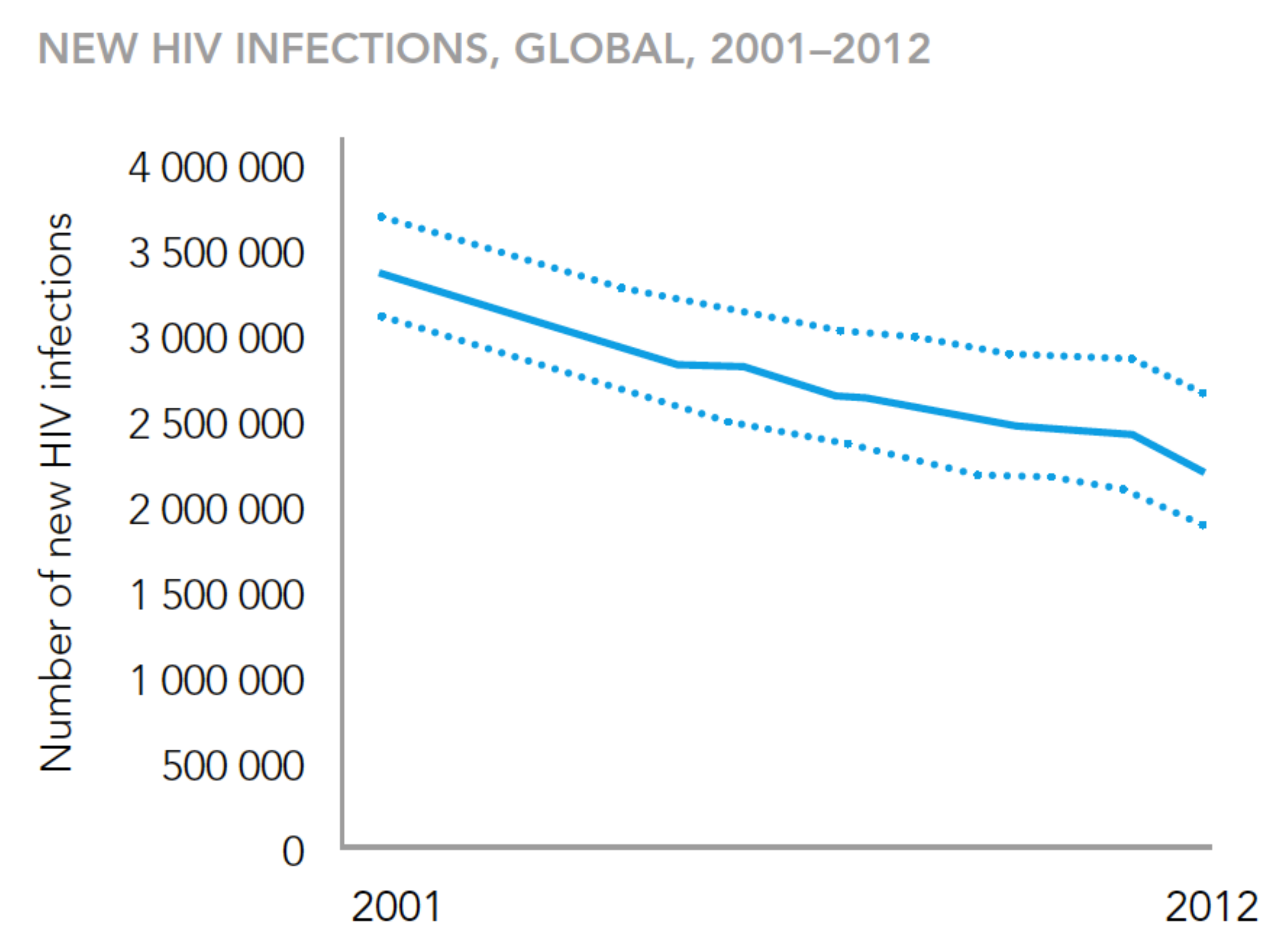} &
\includegraphics[width=0.45\textwidth]{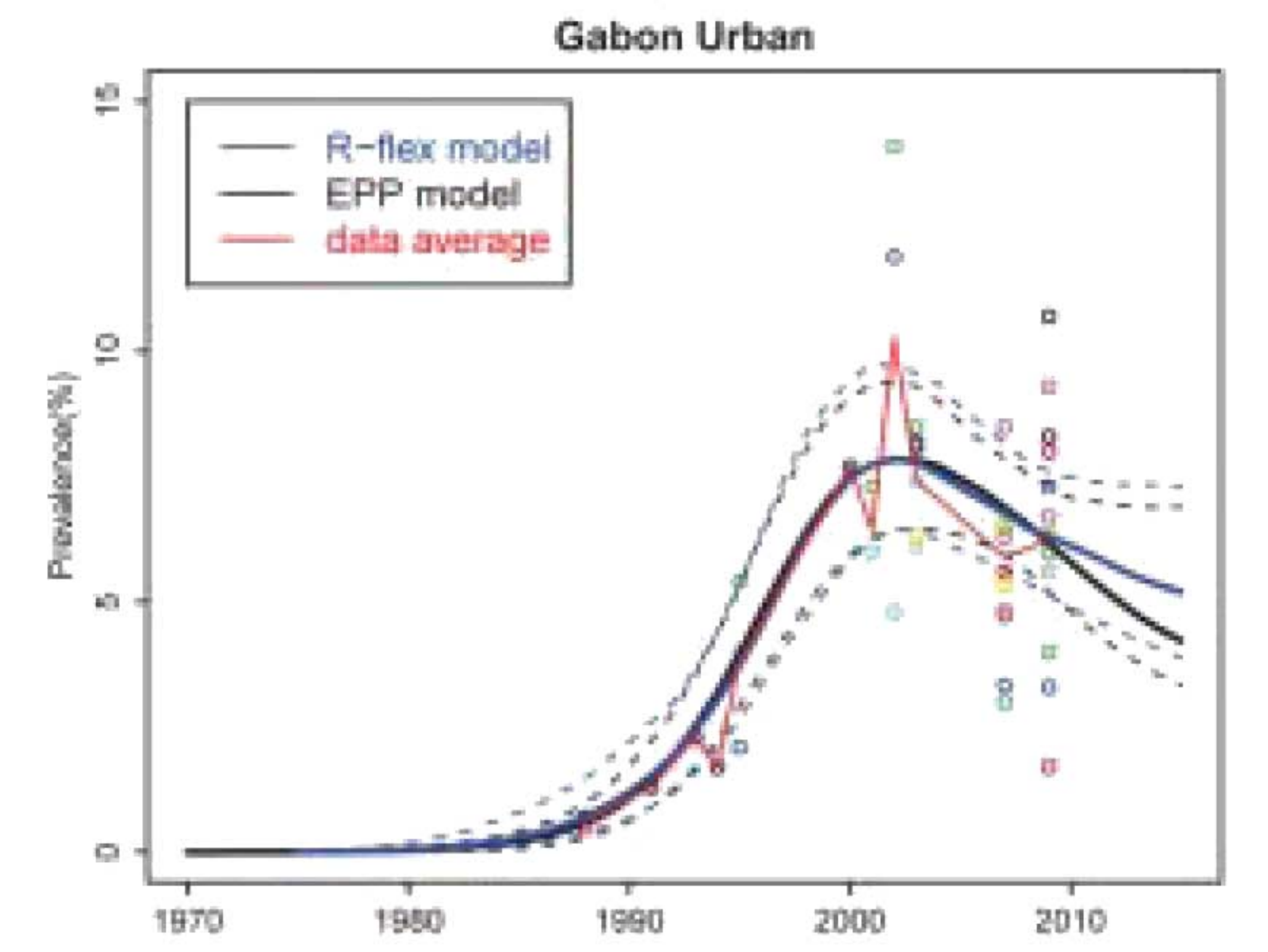} 
\end{tabular}
\end{center}
\caption[Uncertainty Statements in Estimates and
Projections of the HIV/AIDS Epidemic]
{\label{fig-unaids} Uncertainty Statements in Estimates and
Projections of the HIV/AIDS Epidemic.
Left: Estimates of global number of new HIV infections, 2001--2012,
with uncertainty bounds, from UNAIDS' main annual public statement.
{\it Source:} \citet{UNAIDS2013}.
Right: Estimates and projections of HIV prevalence in Gabon, 1970--2015,
based on antenatal clinic data up to 2009; the bands for 2010--2015 
summarize probabilistic projections. The results are shown for 
two different models, EPP and R-flex, and the observed prevalence at
individual clinics is shown by unfilled circles.
{\it Source:} \citet{Bao&2012}. }
\end{figure}

The second goal is to contribute to the basis for the UNAIDS annual reports
\citep{UNAIDS2013}. Uncertainty statements about estimates are routine
in the UNAIDS reports, perhaps because UNAIDS is a newer agency,
established in 1996, by which time it had become the norm
to include uncertainty measures of some kind with estimates of uncertain
quantities.  See Figure \ref{fig-unaids}(a) for an example.
While the uncertainty statements do not feature prominently in the
published report for the broad public, they underlie assessments in the
report such as the following:
\begin{quote}
``The annual number of new HIV infections among adults and adolescents 
decreased by 50\% or more in 26 countries between 2001 and 2012. 
However, other countries are not on track to halve sexual HIV transmission, 
which underscores the importance of intensifying prevention efforts.''
\end{quote}
The phrase ``not on track'' reflects conclusions drawn in part from 
probabilistic projections.

We developed methods for assessing uncertainty about estimates
and projections using Bayesian melding
\citep{Alkema&2007,Brown&2008,Bao&2012}.
One example of the output is shown in Figure \ref{fig-unaids}(b).
Figures such as these typically do not make their way into the most
visible public reports; instead they provide background support
for the conclusions presented in these reports.

There seem to be two main kinds of use for the probabilistic
estimates and projections developed by UNAIDS.
The first is to provide a general assessment of the estimates
and projections and how accurate they are likely to be.

The second kind of use is to assess changes. 
For example, reported HIV prevalence
might increase in a given year, but the question then arises whether
the increase is out of the range of normal expections, perhaps warranting
some new policy intervention. Probabilistic forecasts such as those 
summarized by the uncertainty bands in Figure \ref{fig-unaids}(b) can
be useful in this context. For example, if the new estimated prevalence
is inside the range of the projection (even if there is an increase), 
then there is little evidence
that what is happening is out of what could be expected in the normal
run of things, and the chosen policy could be to continue as before
while monitoring the situation. On the other hand, if the
new estimate is outside the projected range, there may be grounds
for concern and for an intervention.

\section{Probabilistic Population Projections for the United Nations}
\label{sect-ppp}
The United Nations (UN) publishes projections of the populations
of all countries broken down by age and sex, updated every
two years in a publication called the {\it World Population Prospects} (WPP).
It is the only organization to do so. These projections are used
by researchers, international organizations and governments, 
particularly those with less developed statistical systems, 
and researchers. They are used for planning, social and health research,
monitoring development goals, and as inputs to other forecasting models
such as those used for predicting climate change and its impacts
\citep{IPCC2007,Seto&2012}.
They are the de facto standard \citep{LutzSamir2010}.

Like almost all other population projections,
the UN's projections are produced using the standard
cohort-component projection method 
\citep{Whelpton1936,Leslie1945,Preston&2001}.
This is a deterministic method based on an age-structured version
of the basic demographic identity that the number of people
in a country at time $t+1$ is equal to the number at time $t$
plus the number of births, minus the number of deaths, plus the 
number of immigrants, minus the number of emigrants.

The UN projections are based on assumptions about future fertility, 
mortality and international migration rates; given these rates the UN produces 
the ``Medium'' projection, a single
value of each future population number with no statement of uncertainty.
The UN also produces ``Low'' and ``High'' projections using total fertility
rates (the average number of children per woman) that are, respectively,
half a child lower and half a child higher than the 
Medium projections. These are alternative scenarios that have 
no probabilistic interpretation.

Like the UN up to 2008, most national statistical offices, including the 
U.S.~Census Bureau and the U.K.~Office of National Statistics, use assumptions
about future fertility, mortality and migration rates from experts:
either internal experts or panels of outside experts.
Expert knowledge is an essential part of the population projection process,
and experts are generally agreed to be good at assembling and reviewing
the underlying science, as well as assessing the actual forecasts.

However, evidence has been mounting over the past 60 years that experts
in several domains are less good at producing forecasts themselves from scratch.
\citet{Meehl1954} found that very simple statistical models
beat expert human forecasters overall in a range of clinical 
disciplines, and this finding has been replicated in many
subsequent studies \citep{Meehl1986}. 
\citet{OeppenVaupel2002} showed that expert forecasts of life expectancy
at birth, both by leading demographers and forecasting organizations,
had performed poorly over the previous 70 years. 
Forecasters generally tended to project that the future would be like 
the present, and in particular that a limit to life expectancy would 
be reached soon, whereas in fact life expectancy continued to 
increase throughout the period.

\citet{Tetlock2005} evaluated the quality of about 3,000 forecasts
of political events and outcomes by experts, many highly distinguished,
and found their performance to be startlingly poor.
He memorably concluded that many of the experts would have been 
beaten by a ``dart-throwing chimpanzee.''
In a rare counterexample, \citet{MandelBarnes2014}
found that analysts in a Canadian intelligence agency provided 
calibrated forecasts of good quality.

In collaboration with the UN Population Division, we developed new
statistical methods for projecting future fertility and mortality 
rates probabilistically, and translating these into 
probabilistic population projections for all countries
(Alkema et al., 2011; Raftery et al., 2012, 2013; Fosdick and Raftery 2014;
Raftery, Lalic et al., 2014; Raftery, Alkema and Gerland, 2014). 
\nocite{Alkema&2011,Raftery&2012,RafteryChunn&2013,FosdickRaftery2014,RafteryLalic&2014,Raftery&2014StatSci}

An experimental version of the new probabilistic projections was
issued by the UN in November 2012, at http://esa.un.org/unpd/ppp.
This release was accompanied by no fanfare, but the experimental 
probabilistic projections have still had about 10,000 downloads per month.
Official UN probabilistic population projections for all countries
were issued for the first time on the same website on
July 11, 2014 (World Population Day). 

There are other indications of the beginning of a paradigm shift from
deterministic population projections based on expert assumptions 
to probabilistic population projections based on statistical models.
Statistics New Zealand changed its official population
projection method to probabilistic projections in 2012 
\citep{Bryant2014}.

But these releases are recent, and it remains to be seen how and
to what extent ultimate users, such as policy-makers and planners,
make use of them. One possible use is in setting future international goals,
similar to the Millenium Development Goals for 2015 for things
like child and maternal mortality.
It is desirable to set goals that are ambitious but also realistic,
and probabilistic projections could be useful in indicating
what is realistic, suggesting setting goals that are towards the
``good'' end of the probability distribution \citep{Gerland2014}.

A possible use of probabilistic population projections
is in making decisions about policy issues that depend directly
on future population numbers, such as school and hospital infrastructure.
One such decision is whether or not to close schools.
These decisions are often based on deterministic population projections,
which can have a spurious air of certainty.
It is not desirable to close a school unless the probability of 
having to reopen it or find other premises in the future is small
\citep{Louis2012}. 

Even if a deterministic population projection points to school enrollments
declining, there can still be a substantial probability of them
staying essentially constant or even increasing, in which case
closing the school would typically not be a good idea.
Basing such decisions on reasonable upper percentiles of future
school enrollments (such as the 90th percentile), rather than
on a deterministic projection or a predictive mean or median, could be a 
reasonable approach.

\section{Conditional Probabilistic Forecasts: How Many Graduate Students
to Admit?}
\label{sect-gradadm}

Like most U.S.~academic departments with graduate programs,
the Department of Statistics at the University of Washington,
of which I am a faculty member, faces the problem of deciding how many 
potential entering graduate students to make funded offers to for the next 
academic year.  Offers are made in December for entry in the following 
September, nine months later, and are binding on the department.

Entering graduate students are funded by a mix of teaching assistantships,
fellowships and research assistantships. 
There are several major uncertainties to deal with in making this 
decision. The number of research assistantships available depends
on the outcome of faculty research grant applications, which are often 
unknown nine months ahead of time. 
Not all students accept our offers, and we do not know ahead of time 
how many will. We also do not know exactly how many current
students will leave in the next nine months through graduation or dropout.

Up to 2009, departmental practice was to make a number of offers based
on expected numbers of students graduating and grants,
and on an assumed acceptance rate. However, these calculations
were based only on expectations and were not probabilistic,
and also did not incorporate past data in a systematic way.

This often led in practice to too few acceptances relative to the number of
positions available, with the result that teaching assistants for
Statistics courses had to be recruited from among non-Statistics
graduate students. This was undesirable in that statistics teaching was 
not being done by optimally qualified people,  
departmental teaching assistantships were ``lost'' to other departments,
and the pool of future potential research assistants was depleted.
Also, there are currently more jobs available for Ph.D.~statisticians
that graduates, so increasing the number of entering graduate students
is desirable from the labor market point of view as well.
In the five years up to 2009, about four teaching assistantships were being 
``lost'' to the department every year, compared with a 
typical incoming class size of about ten graduate students.

The downside is that if students accept and there is no identified
funding for them, the deparment has to scramble to find funding.
This is difficult but possible within the university, because
many non-Statistics departments have research and teaching needs for 
statistically qualified people that they find it hard to meet from 
within their own pool of students.

In 2010, the departmental faculty decided to base the decision
about the number of students to admit on a probabilistic calculation 
instead of the then current expectation-based approach, 
and I took on the task of developing the appropriate method.
For each possible number of offers, I computed the predictive
probability distribution of the number of TA positions lost to 
the department, as this seemed to be the key quantity for decision-making.
Ideally this would be equal to zero.

With perfect knowledge, the number of TA positions lost conditional
on a given number of offers is equal to
\begin{equation}
Y = T + R_1 + R_2 + G + L + D - C - A,
\label{eq-lostTA}
\end{equation}
where
\begin{eqnarray*}
Y &=& \mbox{ Number of TA positions lost to department } \\
T &=& \mbox{ Number of TA positions available} \\
R_1 &=& \mbox{ Number of RA positions available within the department} \\
R_2 &=& \mbox{ Number of RA positions available outside the department} \\
G &=& \mbox{ Number of students graduating by September} \\
L &=& \mbox{ Number of students dropping program by September} \\
C &=& \mbox{ Number of current students} \\
A &=& \mbox{ Number of acceptances}.
\end{eqnarray*}
$T$ and $C$ are taken as known exactly, but the other quantities in
equation (\ref{eq-lostTA}) are uncertain at the time when the
decision has to be made.

The predictive distributions of $R_1$, $R_2$, $G$, $L$, and $A$ are
derived from past data and elicited information. They are treated
as independent in order to derive a joint distribution.
The predictive distribution of $A$ depends on the number of offers, $O$,
and is modeled as Binomial $(O,\pi)$, where $\pi$ is estimated from
historical data.
The predictive distribution of $R_1$ is obtained by polling
departmental faculty to elicit from each of them a predictive
distribution of the number of research assistantships they will have
available in the next academic year. The distribution of 
$R_1$ is then the distribution of the sum of the numbers from
faculty, obtained by convolving the elicited distributions.
The predictive distributions of $R_2$ and $L$ are based on historical
data on these quantities; empirical rather than model-based distributions
are used. The predictive distribution of $G$ is based on current
information about student progress and is typically quite tight.

The predictive distribution of $Y$, the number of lost TA positions,
which is the primary quantity for decision-making, is then obtained 
by Monte Carlo.
A large number of values of each of $R_1$, $R_2$, $G$, $L$, and $A$
are simulated from their predictive distributions, and the corresponding
simulated values of $Y$ are found from equation (\ref{eq-lostTA}).

\begin{figure}
\begin{center}
\includegraphics[width=\textwidth]{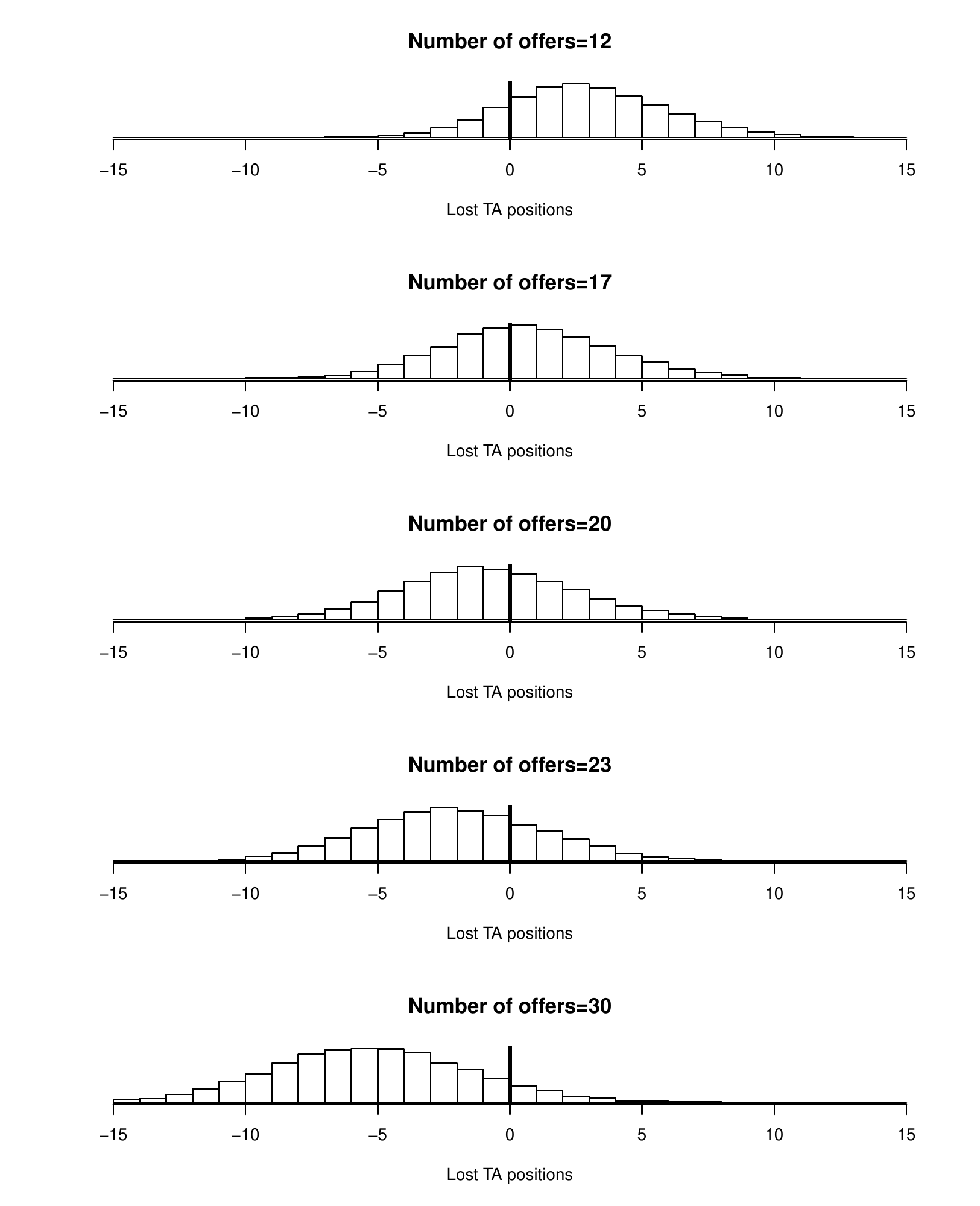}
\end{center}
\caption[Conditional Probabilistic Forecasts of
the Number of Lost Teaching Assistant Positions given Different Numbers
of Graduate Student Offers with Funding]
{\label{fig-GradAdm12} Conditional Probabilistic Forecasts of
the Number of Lost Teaching Assistant Positions given Different Numbers
of Graduate Student Offers with Funding. Negative values indicate the number 
of students that would not be funded from current funding sources.}
\end{figure}

Figure \ref{fig-GradAdm12} shows conditional predictive distributions of 
the number of lost TA positions given several possible number of offers,
and Table \ref{tbl-GradAdm12} shows percentiles of these distributions.
Note that negative numbers correspond to the number of students
that would not be funded with current funding sources.
In these cases, alternative funding sources would be sought,
such as research or teaching assistantships in departments that 
currently fund few or no statistics graduate students.

\begin{table}
\caption[Percentiles of the Predictive Distributions
of the Conditional Probabilistic Forecasts of the Number of Lost Teaching
Assistant Positions given Different Numbers of Graduate Student Offers
with Funding]
{\label{tbl-GradAdm12} Percentiles of the Predictive Distributions
of the Conditional Probabilistic Forecasts of the Number of Lost Teaching 
Assistant Positions given Different Numbers of Graduate Student Offers 
with Funding. Negative values indicate the number of students that would 
not be funded from current funding sources.}
\begin{center}
\begin{tabular}{crrrrrl} \\ \hline
\#Offers & 10\% & 33\% & 50\% & 67\% & 90\% & Description \\ \hline 
12 & 0 & 2 & 3 & 5 & 7 & Very conservative  \\
17 & -3 &  0 & 1 & 2 & 5 & Conservative  \\
20 & -4 & -2 & 0 & 1 & 4 & Break-even  \\
23 & -6 & -3 & -2 & 0 & 3 & Bold  \\
30 & -9 & -6 & -5 & -3 & 0 & Very bold \\ \hline
\end{tabular}
\end{center}
\end{table}

The verbal descriptions in Table \ref{tbl-GradAdm12} characterize
how aggressive a decision is relative to the uncertainty.
For example, 20 offers is the break-even point, because with that number
the department is equally likely to lose TA positions as to have to
seek additional funding sources. Similarly, 23 offers is described as 
``bold'' because there is only one chance in three of losing TA positions,
but a larger chance of having to seek additional positions.

Given these numbers, the then department chair decided to take a ``bold''
stance and make 23 offers. Under the previous system fewer
offers would likely have been made. In the event, the department was able to
fund the students who accepted the offers quite comfortably, so that 
the bold stance turned out well. The previous expectation-based more 
conservative approach could have led to several TA positions being
lost to the department, as in the preceding years. 
The probabilistic approach made it possible
to go beyond the break-even point, and to quantify the risk in so doing,
thus helping the decision-maker to decide how far beyond the
break-even point to go. Given this successful outcome, the 
department decided to continue to use this approach, which has now
been used in four successive years.

The decision to be made in this case involves trading off losses of
different kinds (lost TA positions against the possible need to seek additional
funding sources outside the department, which could be difficult and
stressful). Thus it would seem like a possible candidate for formal
decision analysis, especially given that the decision-makers 
are trained statisticians. Nevertheless, a loss function was not assessed
at any point, and decision theory was not used; the predictive distributions 
by themselves provided enough information to the decision-maker.
After the fact, it seems possible to argue that the decision-maker was
using a loss function under which losing a TA position was twice
as bad as having to find funding for an additional student outside
current sources, but if so this was never explicitly articulated.

It would be possible to improve the statistical model used for generating the
probabilistic forecasts. For example, the students the department
ranks most highly for funding typically are less likely to accept
the offer, because they often have more options. However, the model
assumes that all students with an offer are equally likely to accept it;
it would be possible to relax this assumption.
Also, a second round of offers is sometimes made, depending on initial
responses to the first round of offers. It would be possible to extend the model
to include the second round, about which decisions are currently
made without similar quantitative analysis. 
But overall, the method seems developed enough to provide useful guidance
to the decision-makers, and there has not yet been a strong demand for
further methodological refinement.

\section{Discussion}
I have described five cases in which probabilistic forecasts have been
used with a certain degree of success. These lead me to identify
five types of potential user of probabilistic forecasts
(where the five cases don't map exactly onto the five types of user):

\begin{enumerate}
\item Low Stakes User: This is a user for whom the stakes are low and/or
the losses from over- and underpredicting are similar.
An example might be someone deciding whether to wear a sweater or a
short-sleeved shirt based on temperature; a single ``best''
temperature forecast will often be enough in this case.

\item General Assessor : This is a user for whom the probabilistic
forecast provides a general assessment of the likely quality of the forecast.
The UNAIDS annual report is a possible example.
This is important also for the process of forecast improvement.
The goal of forecast development should be to improve forecast accuracy,
and hence to reduce the uncertainty around the forecast 
\citep{SonejiKing2012}.
It is hard to guide this process without an accurate assessment of
forecast uncertainty.

\item Change Assessor: For this kind of user, the probabilistic forecast 
provides a way of assessing whether a change in some measurement is in line
with expectations, or instead is a source of concern warranting action.
An example might be the probabilistic forecasts of HIV prevalence
produced by UNAIDS, where some changes (including increases) are to be expected,
but larger increases that are ``significant'' would sound an alarm.
One-number forecasts provide no way of making this kind of assessment.

\item Risk Avoider: Here the goal includes keeping the risk of an 
adverse outcome to an acceptable level. The IWC bowhead whale quota
is a good example of this, in which the risk of possible damage to the stock
from aboriginal whaling was to be kept to a low level.
Note that this did lead to a ``one number'' forecast, but the forecast
was not the ``best'' or ``central'' forecast, but rather a lower 
percentile of the predictive distribution, in this case the 5th percentile.

\item Decision Theorist: This user has an explicit loss function and is
able to quantify it.
He or she uses the probabilistic forecast to explicitly minimize 
expected loss, as advocated by formal decision theory.
This did not arise in any of the cases I described, and seems most likely
when the different kinds of possible loss being traded off are on the
same scale, typically money. One example would a wind energy company,
which needs to bid on a contract to supply a given amount of energy,
with specified penalties if the contract is not fulfilled
\citep{Pinson&2007}.
\end{enumerate}

The fact that there are different types of user and use of 
probabilistic forecasts suggests that it is important
for developers of probabilistic forecasts to 
interact with users and consider their goals.
While this may seem obvious, it is often not done.
Interaction can take the form of direct contact (meetings, phone,
email and so on) between developers and users. This can be in the
context of an established scientific advisory committee with
regular meetings and an official membership (as used by the IWC),
or a small less formal reference group with rotating members
(as used by UNAIDS), or expert group meetings, which are effectively
workshops  lasting several days (as used by the UN Population Division). 
If the probabilistic forecasts are delivered
to the general public using a website (as in the case of probabilistic
weather forecasting), the interaction can take the form of a web survey
\citep{JoslynSavelli2010}.

It is important for trust in the forecast that the probabilistic
statements be at least approximately calibrated, so that, for example,
events given predictive probability 80\% happen about 80\% of the time
on average. For the forecast to be useful, it is also important that
forecast intervals be narrow, or sharp, enough to provide a basis for action.
Indeed, \citet{Gneiting&2007} defined the key design principle
of probabilistic forecasting as being to maximize sharpness
subject to calibration, and this has been widely accepted.

The experience I have described suggests that formal decision theory,
much advocated in theory by statisticians and economists,
may have less practical application than sometimes claimed.
One reason may be that people are often not aware of their loss
functions. Another may be that using formal decision theory greatly
increases the cognitive load, in that one's loss function has to be
assessed and then the decision theoretic calculations performed.
One also needs to be careful because in practice people tend to attribute 
different values to equivalent losses and gains, contrary to decision theory,
a finding referred to as ``prospect theory'' 
\citep{KahnemanTversky1979,TverskyKahneman1992}.

Nevertheless, a recent result suggests that the scope of decision
theory may be wider than I have conceded. \citet{Gneiting2011}
showed that if the loss function is generalized piecewise linear
as a function of the quantity being predicted probabilistically,
then the optimal point forecast is a quantile of the predictive distribution.
An important special case of this is when the cost of an overestimate
is a fixed multiple of the cost of an underestimate. 
This will often be at least roughly true, and it may be much easier to
elicit that multiple than the full loss function.
People may be able to say, at least approximately, how much worse
an overestimate is than an underestimate, or vice versa.
This also greatly simplifies the practical use of decision theory,
reducing it to the calculation of a predictive quantile,
so that the cognitive load is little greater than that of probabilistic
forecasting by itself.

One overarching conclusion is that people can use and understand
probabilities and probabilistic forecasts, even if they do not
have advanced training in statistics. 
The cognitive research shows that probabilistic forecasts lead
to better decision-making than deterministic ones, and also to increased
trust in the forecast by users. Experience with probabilistic weather
forecasting and probabilistic population projection websites has shown
that there is considerable public interest in probabilistic forecasts, 
even in the absence of much publicity. 
This suggests that once probabilistic forecasts become
available in a domain, they will be used: ``Build it and they will come.''

\paragraph{Acknowledgements:} This work was supported by the Eunice Kennedy 
Shriver National Institute of Child Health and Development through grants nos. 
R01 HD054511 and R01 HD070936, and by a Science Foundation Ireland 
E.~T.~S.~Walton visitor award, grant reference 11/W.1/I2079.
The author is grateful to Geof Givens, Susan Joslyn, Giampaolo Lanzieri
and Elizabeth Thompson for helpful comments and discussions, 
and to Nial Friel and the School of Mathematical Sciences at 
University College Dublin for hospitality during the preparation of this paper.

\bibliographystyle{chicago}
\bibliography{probforecasting}

\end{document}